\documentclass[12pt,letterpaper]{JHEP3}

\usepackage{slashed}
\usepackage{amsmath}
\usepackage{amssymb}
\usepackage{axodraw}

\newcommand{\ba}{\begin{eqnarray}}
\newcommand{\ea}{\end{eqnarray}}
\newcommand{\rmi}[1]{{\mbox{\scriptsize #1}}}
\newcommand{\rmii}[1]{{\mbox{\tiny #1}}}

\newcommand{\tr}{{\rm tr\,}}
\newcommand{\nn}{\nonumber \\}
\newcommand{\fr}[2]{{\frac{#1}{#2}\,}}

\renewcommand{\(}{\left(}
\renewcommand{\)}{\right)}

\newcommand{\e}{\epsilon}

\newcommand{\tfr}[2]{{\textstyle \frac{#1}{#2}\,}}

\def\openone{\rlap 1\kern 0.22ex 1}

\newcommand{\bv}[1]{\mbox{\textbf{#1}}}

\preprint{}
\title
    {
    \boldmath
    Heavy flavor diffusion in weakly coupled ${\mathcal N} =4$ Super Yang-Mills theory
    }
\author
    {%
    P.~M.~Chesler\footnote{\tt pchesler@u.washington.edu}\;
    and A.~Vuorinen\footnote{\tt vuorinen@phys.washington.edu}
    \\Department of Physics, University of Washington, Seattle, WA 98195--1560
    }%

\abstract {We use perturbation theory to compute the diffusion
coefficient of a heavy quark or scalar moving in ${\mathcal N} =4$
SU($N_c$) Super Yang-Mills plasma to leading order in the coupling
and the ratio $T/M\ll 1$. The result is compared both to recent
strong coupling calculations in the same theory and to the
corresponding weak coupling result in QCD. Finally, we present a compact and simple formulation of the
Lagrangian of our theory, $\mathcal N=4$ SYM coupled to a massive fundamental $\mathcal
N=2$ hypermultiplet, which is well-suited for weak coupling expansions.}


\begin{document}


\section{Introduction}

An important and interesting challenge facing theorists investigating heavy ion physics is to predict the rate of energy loss of a
heavy quark moving through a quark gluon plasma (QGP), which is a quantity of direct experimental relevance \cite{exp1,exp2,exp3}. For weak coupling and
ultrarelativistic quarks,
$\gamma v \gg 1/g\gg 1$, the dominant mechanism for this is gluon bremsstrahlung, while for the experimentally equally important region $\gamma v \lesssim 1$,
the energy loss occurs through elastic collisions with light plasma constituents. Both of these cases have been extensively
studied \cite{ultra1,ultra2,ultra3,ultra4,mt}, but all existing calculations share the fundamental shortcoming that they assume the plasma to be weakly coupled,
which need not be the case in the energy range of interest for instance for RHIC physics.

While one has grown to rely upon lattice QCD as the source of direct information on the strong coupling regime of various static observables, it is at present
still a relatively inefficient tool in the description of real-time phenomena (for recent advances, see \textit{e.g.} Ref.~\cite{berg}).
A lot of attention has therefore been turned towards addressing the question of the energy loss rate of a quark moving in a strongly coupled non-Abelian plasma
in an entirely different framework, the ${\mathcal N} =4$ supersymmetric Yang-Mills (SYM) theory with gauge group SU($N_c$). There, one has the unique
opportunity of being able to access analytically the strong coupling limit of the quantum field theory
(in its large $N_c$ limit) via the famous AdS/CFT conjecture that relates the theory to dual type IIB supergravity on an $AdS_5\times S^5$ background \cite{malda}.
The energy loss calculation then reduces to studying classical string dynamics in the $AdS_5$ background, which has yielded useful analytical results for the heavy
(and light) quark energy loss parameters in the strong coupling limit (see Refs.~\cite{hky,ct,mit,ch,cg,gubs,mit2} and references therein).

While the development of non-equilibrium lattice field theory methods as well as the search for dual string theories of more QCD-like theories continue, it is
worthwhile to first ask the more modest question of what kind of qualitative, or even quantitative, insight the QCD
community can draw from the existing SYM calculations. An obvious way of addressing this is to perform similar weak coupling calculations in the SYM theory
that have been carried out in the QCD context and compare the results on one hand to the the weak coupling limit of QCD and on the other hand to the strong
coupling limit of the $\mathcal N=4$ SYM. This can be expected to yield valuable information on the similarities and differences of the two theories and to furthermore
indicate, to which extent one can simply extrapolate the existing weak coupling results in QCD to strong coupling. In the present paper, we aim to do exactly
this by investigating the simplest observable related to the energy loss of a non-relativistic heavy quark in the SYM theory, its diffusion coefficient, in the
weak coupling regime. The calculation is to a large extent parallel to the corresponding QCD computation of Moore and Teaney \cite{mt} and generalizes many of
its results to the SYM case.

Our paper is organized as follows. In Section 2, we briefly introduce $\mathcal N = 4$ SYM and write down its Lagrangian in a form useful for weak
coupling expansions, after which we discuss when and how one may use semi-classical kinetic
theory techniques to obtain the diffusion coefficient of a heavy particle in a quantum field theory. In Section 3, we then go through the necessary calculations,
after which we display our main result for the heavy quark diffusion coefficient, $D_\rmi{Q}$, that is to be further analyzed and discussed in Section 4.
In Section 5, we finally draw conclusions and outline some future work to be carried out through weak coupling kinetic theory calculations in the SYM theory.
The Appendices contain further computational details, such as a relatively detailed derivation of our Lagrangian as well as a discussion of how the necessary
scattering amplitudes and integrals were evaluated.

Throughout the paper, we will for convenience use notation adopted from Refs.~\cite{mt,wein}. This implies working with the Minkowski space metric $-+++$ and denoting
four-vectors by capital letters $P, Q$, three-vectors by bold ones \textbf{p, q} and the absolute values of the latter by $p, q$. The Dirac gamma matrices
are defined so that $\gamma_0$ is anti-hermitian while the $\gamma_i$ are hermitian, and consistently with this we have
$\gamma_5\equiv i\gamma_0\gamma_1\gamma_2\gamma_3$ and $\beta\equiv i\gamma_0$. The gauge is fixed by choosing to work in the Coulomb gauge.

For the most part, we will work with four-component Majorana spinors that can be written in the special form
\ba
\psi&=&\binom{e\zeta^*}{\zeta},
\ea
where $e\equiv i\sigma_2$ and $\zeta$ denotes a two-component Weyl fermion. For the Majorana spinors,
\ba
\label{majorana}
\bar{\psi}&\equiv&\psi^\dagger\beta=\psi^T\e\gamma_5,
\ea
with $\e$ being the 4$\times$4 matrix
\begin{displaymath}
\e=\left(
\begin{array}{cc}
e\;&\;0\\
0\;&\;e
\end{array}
\right).
\end{displaymath}

\section{The setup}

\subsection{${\mathcal N} =4$ SYM with massive quarks}
The theory we consider is ${\mathcal N} =4$ Super Yang-Mills coupled to a ${\mathcal N} =2$ heavy fundamental hypermultiplet. The field content of
${\mathcal N} =4$ SYM consists of a gauge field $A_{\mu}$, four Majorana fermions $\psi_i$ and three complex scalars $\phi_p$, while
the additional $\mathcal N=2$ multiplet is composed of two heavy scalars $\Phi_n$ and a Dirac fermion $\omega$. All $\mathcal N=4$ fields transform under
the adjoint representation of the gauge group SU($N_c$) and are therefore traceless hermitian $N_c\times N_c$ matrices, while the $\mathcal N=2$ sector consists of
$N_c$ component vectors in color space transforming under the fundamental representation.

Following Ref.~\cite{yy}, we define $\phi_p = 1/\sqrt{2} \(X_p + i Y_p \)$, with $X_p$ and $Y_p$ hermitian, which allows us to write our Lagrangian in the form
\ba
\label{lagrangian}
\mathcal L = \mathcal L_0 + \mathcal L_1 + \mathcal L_2,
\ea
with
\vspace{-1.2cm}

{\setlength{\baselineskip}{1.6\baselineskip}
\ba
\label{l0}
\mathcal L_0&=&  -\tr \Big \{\frac{1}{2} F_{\mu \nu} F^{\mu \nu}+ \bar \psi_i \slashed{D} \psi_i +\(D X_{p}\)^{2}+\(D Y_{p}\)^{2} \Big \}  \nn
&-&\Phi_n^{\dagger} (-D^{2}+M^2) \Phi_n -\bar \omega (\slashed{D} +M)\omega,\\
\label{l3}
\mathcal L_1/g &=& \tr \Big \{-i \bar \psi_i \alpha_{ij}^{p} [X_p,\psi_j] + \bar \psi_i \gamma_5 \beta^{p} _{ij}[Y_p, \psi_j] \Big \} -\bar \omega
\left ( Y_1 - i \gamma_5 X_1 \right ) \omega  \nn
&+&2 \sqrt{2} \,\text{Im} \Big( - \bar \omega P_{+} \psi_1 \Phi_1 - \Phi_2^{\dagger} \bar \psi_1 P_{+} \omega +  \Phi_1^{\dagger} \bar \psi_2 P_{+} \omega
- \bar \omega P_{+} \psi_2 \Phi_2 \Big) \nn
&-&2 M \Phi^{\dagger}_{n} Y_{1} \Phi_{n},
\ea
\ba
\label{l4}
\mathcal L_2/g^2 &=&-\frac{1}{2} \tr \( i [\chi_A, \chi_B]\)^2+(-1)^{n}  \Phi_n^{\dagger} \( [\phi_2,\phi_2^{\dagger}] +
[\phi_3,\phi_3^{\dagger}] \) \Phi_n \nn
&-& 4 \text{Re} \( \Phi_1^{\dagger} [\phi_2,\phi_3] \Phi_2\)-\tfr{1}{2} \big | (-1)^{n} \Phi_{n}^{\dagger} t_a  \Phi_{n} \big |^{2}
-2  \big | \Phi_2^{\dagger} t_a \Phi_1 \big |^{2} \nn
&-& \Phi^{\dagger}_{n} \{\phi_1,\phi_1^{\dagger} \} \Phi_{n}.
\ea
\par}
\noindent Here, $D$ denotes covariant derivatives in the appropriate representations of SU($N_c$), $\chi \equiv (X_1,Y_1,X_2,Y_2,X_3,Y_3)$ and a sum over repeated
indices is implied. The matrices $\alpha^p$ and $\beta^p$ are given by
\begin{subequations}
\label{coefficients}
\begin{align}
    \alpha^1&=
    \begin{pmatrix}
        i\sigma_2 & 0    \\
        0 & i\sigma_2
    \end{pmatrix}   ,\;\;\;\;
    \alpha^2=
    \begin{pmatrix}
        0 & -\sigma_1   \\
        \sigma_1 & 0
    \end{pmatrix}   ,\;\;\;\;\;\;
    \alpha^3=
    \begin{pmatrix}
        0 & \sigma_3   \\
         -\sigma_3 & 0
    \end{pmatrix}   ,\;
                \\
    \beta^1&=
    \begin{pmatrix}
        -i\sigma_2 & 0   \\
        0 & i\sigma_2
    \end{pmatrix}   ,\;
    \beta^2=
    \begin{pmatrix}
        0 & -i\sigma_2    \\
        -i\sigma_2 & 0
    \end{pmatrix}   ,\;
    \beta^3=
    \begin{pmatrix}
        0 & \sigma_0  \\
         -\sigma_0 & 0
    \end{pmatrix}   \, ,
\end{align}
\end{subequations}

\noindent and they satisfy the algebra
\ba
\{\alpha^p,\alpha^q\} &=& \{\beta^p,\beta^q\} = -2 \delta^{pq} ,\nn
\big[\alpha^p,\beta^q\big] &=& 0.
\ea
\noindent For more details on the derivation of this Lagrangian, see Appendix \ref{deriveL}.


From the form of the Lagrangian it is clear that neither the heavy fermion nor the heavy scalar number is independently conserved, as Eq.~(\ref{l3}) fails to be
invariant under the separate global $U(1)$ transformations
\ba
\label{trans}
\Phi_i &\rightarrow& U_{\Phi} \Phi_i, \nn  \omega &\rightarrow& U_{\omega} \omega ,
\ea
where $U_{\Phi}$ and $U_{\omega}$ are independent phases. This implies that the diffusion coefficients for heavy quarks and heavy scalars are in general not
independently
well-defined. However, if $U_{\Phi}=U_{\omega}$, the transformation of Eq.~(\ref{trans}) does leave the Lagrangian invariant, which means that this combined
transformation gives rise to a conserved heavy flavor current that includes both fermions and scalars. It is the diffusion of this heavy flavor density that we are
interested in.

\subsection{Diffusion of a heavy non-relativistic particle}
Following the approach of Ref.~\cite{mt}, let us consider the kinematics of a heavy particle immersed in weakly coupled $\mathcal N=4$ SYM plasma at temperature $T$.
We assume that the particle
is in thermal equilibrium with the plasma and that its mass $M \gg T$, in which case the typical energy of all types of excitations is $E\sim T$ and the typical
momentum of the
heavy particle is $p \sim \sqrt{MT}$. At weak coupling, the dominant scattering processes for the heavy particles are $2\leftrightarrow2$ elastic collisions with light
plasma constituents,
in which the typical momentum exchanged is $q\sim T$ and the typical change in the heavy particle velocity is $\delta v \sim T/M \ll 1$. It thus takes many collisions
for the velocity to change significantly, and consequently the collisions may be treated
as uncorrelated events, in which the heavy particles receive random kicks from the medium. In addition, the mean free path of the heavy particle
$\lambda_\rmii{MFP} \sim\( \frac{M}{T}\)^2 \frac{1}{g^4 T \log \frac{1}{g}}$ is parametrically large in comparison with its de Broglie wavelength, thus
allowing one to use semiclassical methods in the treatment of its dynamics.

Let us then look at the trajectory of a heavy particle that starts from the origin at $t=0$ and denote its position at time $t$ by $\bv x(t)$. Under the
assumption that collisions with the medium force the heavy particle to undergo a random walk, the diffusion coefficient $D$ is defined by
\ba
\langle x^{2}(t) \rangle = 6 D t.
\label{Ddef}
\ea
Denoting the random force exerted on the particle by the medium by $\mathbf{\xi}(t)$, the above assumptions imply that the
dynamics of the particle follow from the Langevin equation (see \textit{e.g.}~Ref.~\cite{reif})
\ba
\frac{d p_i}{d t} = \xi_i(t) - \mu p_i,
\ea
where $\mu$ is the momentum drag coefficient. As collisions with the light particles are uncorrelated, we furthermore have
\ba
\label{Diff}
\langle \xi_i(t) \xi_j(t') \rangle = \kappa \delta_{ij} \delta(t-t'),
\ea
where $3 \kappa$ is the mean squared momentum transfer per unit time. Using the equilibrium relation $\langle p^{2} \rangle = 3 M T$ as well as the solution
to the above differential equation, it is then easy to show that \cite{mt}
\ba
\label{eta}
\mu = \frac{\kappa}{2 M T},\quad
\langle x^2(t) \rangle = \frac{6 T t}{M \mu},
\label{Diff2}
\ea
from which it follows that the heavy particle diffusion coefficient is given by
\ba
\label{diffconst}
D = \frac{2 T^2}{\kappa}.
\ea

The above result relating the diffusion coefficient to $\kappa$ proves highly useful for our purposes, as in the
semiclassical regime where kinetic theory is valid we may
immediately write down an expression for the latter in terms of the scattering amplitudes of the quantum theory. Denoting
the heavy particles by $H$, the light particles by $\ell$ and the Bose and Fermi distribution functions by $n_b(k)$ and $n_f(k)$, respectively, the
mean squared momentum transfer per unit time is given by \cite{mt,amy1}
\ba
\label{fullkappa}
3\kappa &=&\fr{1}{16(2\pi)^5M^2}\int\fr{d^3\bv{k}\,d^3\bv{k}'d^3\bv{p}'}{k_0 k'_0}\delta^3(\bv{p}-\bv{p}'+\bv{k}'-\bv{k})\delta(k-k')\nn
&\times& \sum_{H \ell, H' \ell'} \Big\{ \left |\mathcal M_{H \ell \rightarrow H' \ell'} \right |^2 n_{\ell}(k)(1\pm n_{\ell'}(k'))\Big\}.
\ea
Here, $\left |\mathcal M_{H \ell \rightarrow H' \ell'} \right |^2$ stands for the scattering amplitudes squared --- summed over all internal degrees of freedom of
the light particles and the final state heavy particle and averaged over those of the initial heavy particle (including the flavors)
--- for the process\footnote{The field theory also allows for heavy particles to scatter
off of other heavy particles, but their contribution to the integral of $\kappa$ is suppressed by an exponential of $M/T$.}
$H \ell \rightarrow H' \ell'$. The plus sign is taken when $\ell'=b$ represents a final state boson and the minus sign when
$\ell'=f$ represents a final state fermion.

\section{Calculations and results}
\label{calcs}

In the non-relativistic limit where $M\gg T$, the number of interaction terms in the Lagrangian relevant for the scattering of massive particles can be
greatly reduced, as there exists a hierarchy in the $M$ dependence of the various scattering amplitudes. First of all, the amplitude for any tree level process that
contains an intermediate heavy particle will be suppressed by an inverse power of $M$ relative to those with an intermediate light particle.
Second, in the non-relativistic limit each external heavy fermion\footnote{We use the convention $\bar u_{s}(\bv p) u_{s'}(\bv p) = 2 M \delta_{ss'}$ in the normalization of the
heavy spinors.} will
introduce a factor of $\sqrt{M}$ to the amplitudes, and each heavy scalar/gluon vertex (with momentum $P \sim (M, \bv 0)$ flowing
through it) as well as each heavy scalar/light scalar vertex will introduce an explicit factor of $M$. Therefore, at leading order in $g$
the diagrams proportional to the highest power of $M$ are those,
in which a heavy fermion or scalar scatters
elastically off of a light plasma constituent via the exchange of a light intermediate boson. These processes are depicted in Fig.~1.a-b, while some examples of
processes whose amplitudes are suppressed by positive powers of $T/M$ are shown in Fig.~1.c. The latter include inelastic processes, diagrams with heavy intermediate
lines and graphs containing a four-scalar vertex that is independent of $M$.
\begin{FIGURE}[ht]
{ \centerline{\def\epsfsize#1#2{0.9#1}\epsfbox{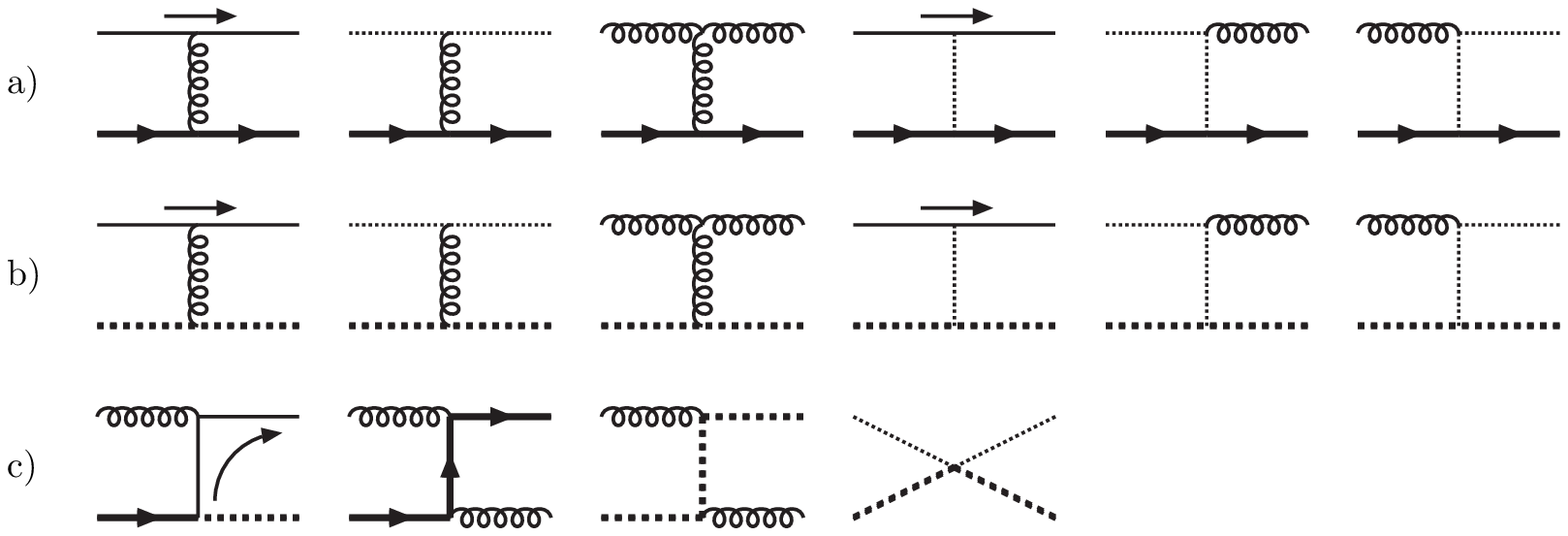}}
\caption[a] {a) The lowest order tree-level processes contributing to
the heavy quark diffusion coefficient, the amplitudes of which come with the maximal
power of $M$. b) The corresponding diagrams relevant for the heavy
scalar diffusion coefficient. c) Examples of processes contributing to
the heavy flavor diffusion coefficient but suppressed by powers of
$T/M$. The solid and dotted bold lines correspond to the heavy
quarks and scalars, respectively, the solid and dotted light lines
to the corresponding massless fields, and the wavy lines to gluons.
The arrows drawn adjacent to the light quark lines indicate the
(arbitrary) direction of the flow of the Majorana fermion number
(see \textit{e.g.}~Ref.~\cite{majo}). \label {fig1} } }
\end{FIGURE}

Finally, we note that in the non-relativistic limit we may neglect the coupling of $X_p$ to heavy fermions, which follows
from the fact that $\bar \omega \gamma_5 \omega$ is parity odd. In the non-relativistic limit, the scattering of fermions via
scalar exchange is independent of spin and thus conserves the parity of the fermions, implying that the heavy fermion scattering amplitudes containing an exchanged
$X_p$ must be suppressed by inverse powers of $M$. In summary, to obtain the desired heavy flavor diffusion coefficient to leading order in $T/M$ and $g$,
we may neglect $\mathcal L_2$ entirely and replace $\mathcal L_1$ by the effective interaction Lagrangian
\ba
\label{l3eff}
\mathcal L^\rmi{eff}_1/g &=& \tr\( \bar \psi_i \gamma_5 \beta^{1} _{ij}[Y_1, \psi_j] \)
-\bar \omega Y_1 \omega-2 M \Phi^{\dagger}_{n} Y_{1} \Phi_{n}.
\ea

After neglecting the species changing Yukawa terms from $\mathcal
L_1$, the Lagrangian becomes invariant under the separate $U(1)$
transformations of Eq.~(\ref{trans}). It then follows that to
leading order in $M$, the heavy fermion and scalar currents are
independently conserved, and therefore the corresponding fermion and
scalar diffusion coefficients $D_\rmi{Q}$ and $D_\rmi{S}$ can be
independently defined. For these currents, the mean squared momentum
transfer per unit time is given by
\ba
\label{fullkappaH}
3\kappa_\rmii{H} &=&\fr{1}{16(2\pi)^5M^2}\int\fr{d^3\bv{k}\,d^3\bv{k}'d^3\bv{p}'}{k_0 k'_0}\delta^3(\bv{p}-\bv{p}'+\bv{k}'-\bv{k})\delta(k-k')\nn
&\times& \sum_{ \ell, \ell'} \Big\{ \left |\mathcal M_{H \ell \rightarrow H \ell'} \right |^2 n_{\ell}(k)(1\pm n_{\ell'}(k'))\Big\},
\ea
where H stands for either Q or S, and the corresponding diffusion coefficients read
\ba
\label{diffconstH}
D_\rmi{H}  =  \frac{2 T^2}{\kappa_\rmii{H}}.
\ea

Even though $D_\rmi{Q}$ and $D_\rmi{S}$ are independently well defined, a quick inspection of the
forms of $\mathcal L_0$ and $\mathcal L^\rmi{eff}_1$ reveals that their values are in fact equal. To see this, note that if the
momentum exchanged in a collision is $\bv q= \bv p' -\bv p \sim T$, where $\bv p$ and $\bv p'$ are the momenta of the incoming and outgoing heavy quarks,
respectively, then the spinors $u_s(\bv p)$ corresponding to the external legs satisfy (up to $\mathcal O (T/M)$ corrections)
\ba
\bar u_s(\bv p) u_{s'}(\bv p') &\approx& 2 M \delta_{ss'}, \nn
\bar u_s(\bv p) \gamma_\mu u_{s'}(\bv p') &\approx& -2 i M \delta_{ss'} \delta_{\mu 0},
\ea
implying that the contributions of the heavy fermions to scattering amplitudes are spin independent.
To obtain the tree level scattering amplitudes that we are interested in
(and that are not sensitive to particle statistics), we may therefore replace the heavy quark by a complex scalar field $\Sigma$, with the factors of $\pm 2 M$ added
explicitly to the corresponding vertices. The couplings of $\Sigma$ to the light fields are then identical to those of $\Phi_n$, and therefore the corresponding
scattering amplitudes agree as well. In what follows, we will exploit this symmetry and only consider the heavy fermion diffusion coefficient.

Before we can proceed to the actual computation of $D_\rmi{Q}$, we must still deal with the fact that the expression for $\kappa_\rmii{Q}$ given in
Eq.~(\ref{fullkappaH}) is infrared sensitive, which can be seen by noting that if one were to use bare propagators for the exchanged bosons
in the scattering amplitudes squared, the resulting integrals in Eq.~(\ref{fullkappaH}) would diverge
in the $q\rightarrow 0$ limit. This can be attributed to the long range potentials
associated with the exchange of the massless bosons which, however, are modified by the interactions with the plasma that cut the divergences off
at the scale $gT$. Taking the interactions into account, the IR problem is naturally dealt with by including self energy corrections to the
corresponding propagators. As kinematics furthermore require that the energy exchanged in a collision be suppressed relative to the spatial momentum
by a factor of $\sqrt{T/M}$, we note that the appropriate self energy corrections are those due to static thermal screening. We can therefore simply
add static screening masses to the propagators in question, and taking into account the fact that only the temporal gluon propagator enters the calculations, obtain as
the required resummed propagators
\ba
D^{00}_{ab}(\bv p)&=&-\fr{1}{p^2+m_\rmii{D}^2}\delta_{ab}, \label{md} \\
G^{p}_{ab}(\bv p)&=&\fr{1}{p^2+m_\rmii{S}^2}\delta_{ab} \label{ms}
\ea
for $A_0$ and $\phi_p$, respectively.

The squared scattering amplitudes for the processes shown in Fig.~\ref{fig1}.a are computed in Appendix \ref{melemnts}. Denoting heavy quarks by $Q$,
light fermions, scalars and gluons by $f$, $s$ and $g$, respectively, and initial and final light particle three-momenta by $\bv k$ and $\bv k'$, the
results read
\begin{subequations}
\label{scatter}
\ba
    \left |\mathcal M_{Qf\rightarrow Qf} \right |^2 &=& 32 g^4 d_A M^2 k^2 (1+\cos \theta) \frac{1}{(q^2 + m_\rmii{D}^2)^2} \nn
    &+&32 g^4 d_A M^2 k^2 (1-\cos \theta) \frac{1}{(q^2 + m_\rmii{S}^2)^2}, \\
    \left |\mathcal M_{Qs\rightarrow Qs} \right |^2 &=& 48 g^4 d_A M^2 k^2 \frac{1}{(q^2 + m_\rmii{D}^2)^2},\\
    \left |\mathcal M_{Qg\rightarrow Qg} \right |^2 &=& 8 g^4 d_A M^2 k^2 (1+\cos^{2}\theta) \frac{1}{(q^2 + m_\rmii{D}^2)^2}, \\
    \left |\mathcal M_{Qs\rightarrow Qg} \right |^2 &=& 8 g^4 d_A M^2 k^2 \sin^2\theta\, \frac{1}{(q^2 + m_\rmii{S}^2)^2},\\
    \left |\mathcal M_{Qg\rightarrow Qs} \right |^2 &=& 8 g^4 d_A M^2 k^2 \sin^2\theta\, \frac{1}{(q^2 + m_\rmii{S}^2)^2},
\ea
\end{subequations}
where $d_A \equiv N_c^2 - 1$ and $\theta$ is the angle between $\bv k$ and $\bv k'$.
These expressions have been summed over all internal degrees of freedom of the light particles as well as over those of the final state heavy quark, and averaged
over those of the initial state heavy quark.

Appendix B shows how to evaluate the integrals appearing in Eq.~(\ref{fullkappaH}). At weak coupling, where one may use the leading order results
(see \textit{e.g.}~Ref.~\cite{yy})
\ba
\label{debyemasses1}
m_\rmii{D}^2 &=& 2 g^2 N_c T^2, \\
\label{debyemasses2}
m_\rmii{S}^2 &=& g^2 N_c T^2,
\ea
consistency in the weak coupling expansion requires that the terms in the integrals proportional to positive powers of $m_\rmii{D}/T$ or $m_\rmii{S}/T$ be neglected,
which allows us to carry out the integrations analytically. Using the relation of Eq.~(\ref{diffconstH}) between $\kappa_\rmii{Q}$ and $D_\rmi{Q}$, we then obtain as our
main result
\ba
\label{results}
D_\rmi{Q}=\fr{12\pi}{d_Ag^4T}\bigg\{\log \fr{2T}{m_\rmii{D}}+\fr{13}{12}-\gamma_E +\fr{1}{3}\log 2+\fr{\zeta'(2)}{\zeta(2)}\bigg\}^{-1},
\ea
to which the leading corrections come in at relative order $O(g)$. This result is independent of the scalar screening mass, which
follows from the fact that as $\cos \theta = 1-\fr{q^2}{2 k^2}$, every term in Eqs.~(\ref{scatter}a)--(\ref{scatter}e) containing $m_\rmii{S}$ is infrared safe and
therefore does not diverge in the limit $m_\rmii{S}\rightarrow 0$. Also, one should take note of the fact that due to the equality of the heavy quark and heavy scalar
diffusion coefficients, the more general heavy flavor diffusion coefficient $D$, given by the average of the two, coincides with Eq.~(\ref{results}) as well. To
better facilitate a comparison with the strong coupling limit of the theory --- in which only the heavy flavor diffusion coefficient is \textit{a priori}
well-defined --- we will in the following sections refer only to the latter quantity also in the weak coupling context.

\section{Discussion}

Having obtained an expression for the heavy quark (flavor) diffusion coefficient in weakly coupled $\mathcal N=4$ SYM theory, it is interesting to analyze it and to
compare it on one hand to the strong coupling result of Herzog \textit{et al.}~and others \cite{hky,ct} and on the other hand to the corresponding weak coupling
calculation in QCD by Moore and Teaney \cite{mt}. An immediate observation from both the expanded result of Eq.~(\ref{results}) and the integral of Eq.~(\ref{fullkappaH})
is that when written in terms of the 't Hooft coupling $\lambda = g^2 N_c$, the only explicit dependence on $N_c$ in the results comes from an overall factor of
$(1-1/N_c^2)^{-1}$. Keeping in mind that
this multiplicative factor can be reintroduced to the results at any later time, we shall in what follows, unless explicitly stated otherwise, consider the large
$N_c$ limit and set $(1-1/N_c^2)^{-1}=1$.

In order to inspect the domain of validity of the small $m_\rmii{D}/T$ and $m_\rmii{S}/T$ expansions, we plot in Fig.~\ref{pot2} our result for $1/(DT)$
obtained both with and without the expansion, with the curve for the latter originating from a numerical evaluation of the integral in Eq.~(\ref{fullkappaH}), in which the weak coupling
expressions for the screening masses given in Eqs.~(\ref{debyemasses1}) and (\ref{debyemasses2}) are used. We observe that the two curves begin to differ
significantly at $\lambda\sim 3/4$, and that the expanded result for $D$ in fact starts to diverge when $\lambda\gtrsim 2$. This unphysical behavior signals
the breakdown of the leading order weak coupling expansion for $D$ and consequently implies that higher order corrections must be taken into account
to gain even qualitative information about the intermediate coupling regime. Not performing the small screening mass expansion amounts to doing a partial resummation
of our results, where some higher order contributions are included, but others, such as those coming from additional processes like bremsstrahlung or from corrections
to the scattering amplitudes or screening masses, are neglected. While it is not \textit{a priori} obvious that this is sufficient to gain quantitative information about the
behavior of $D$ at larger coupling, it is clear from Fig.~\ref{pot2} that including these corrections improves the qualitative behavior of the diffusion coefficient,
as the unphysical divergence of $D$ at $\lambda \approx 3$ is removed.
In what follows, we will therefore not use the small screening mass expansion, but instead evaluate the integrals of Eq.~(\ref{fullkappaH}) numerically.
\begin{FIGURE}[t]
{
\centerline{\epsfxsize=10.0cm\epsfysize=6.5cm\epsfbox{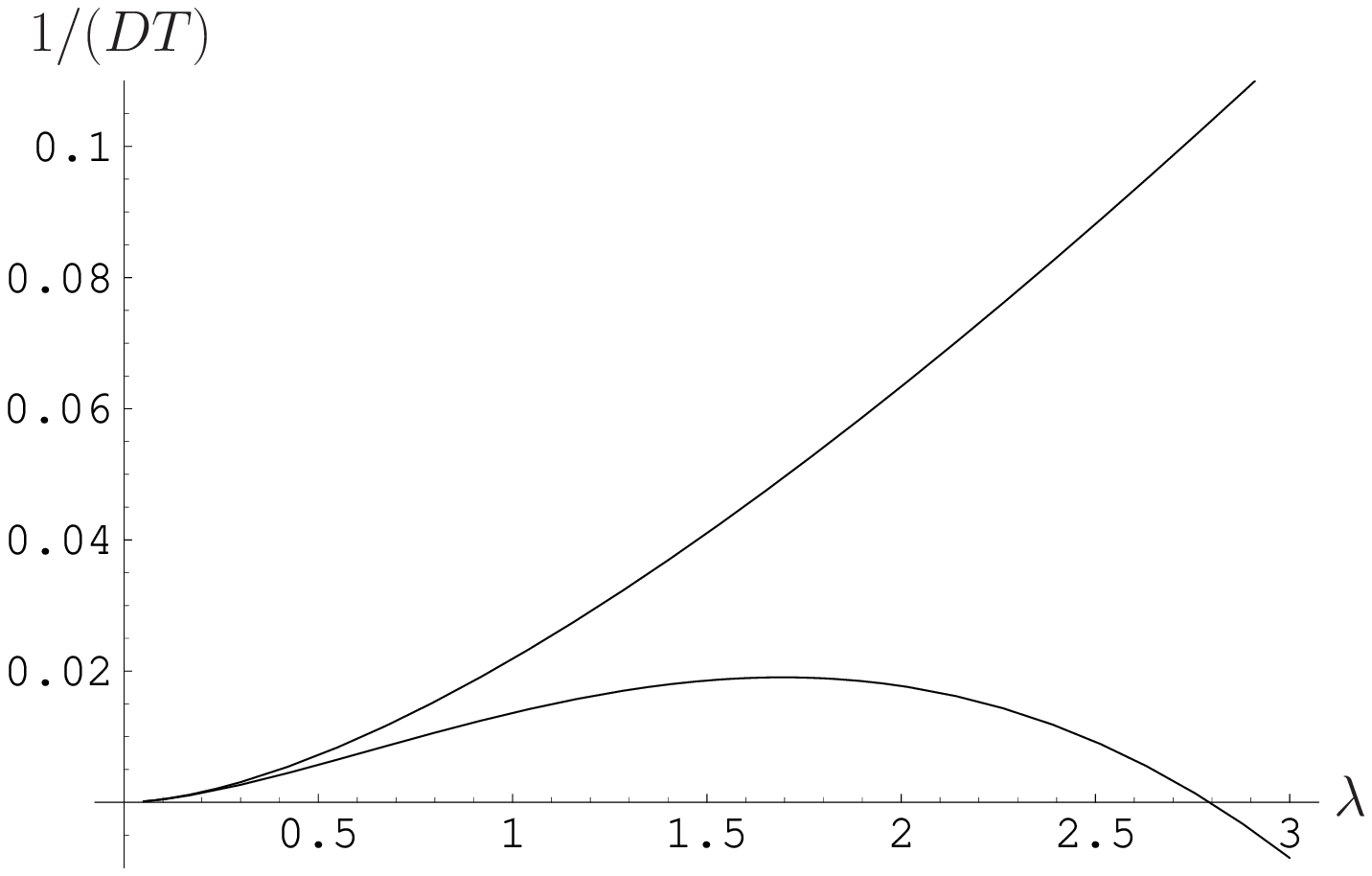}}
\caption[a]%
{Plots of our two weak coupling results for $1/(DT)$: the expanded version taken from Eq.~(\ref{results}) (lower curve) and the one obtained via a numerical evaluation of the
integral in Eq.~(\ref{fullkappaH}) (upper curve). The large $N_c$ limit has been taken here, and the weak coupling expressions for the screening masses have been used.
\label{pot2}}
}
\end{FIGURE}

In Fig.~\ref{fig:pot}.a, we investigate the behavior of our result at stronger coupling by plotting $1/(DT)$ as a function of $\lambda$, with the screening
masses still given by their leading order weak coupling expressions. As $\lambda$ increases, an increasingly important source of ambiguity in this plot comes from neglecting
the NLO corrections to the screening masses, which are expected to be sizable already at $\lambda \sim 1$.\footnote{At these couplings, the leading order weak coupling
results for the screening masses yield values $\gtrsim T$. For consistency, one should therefore use resummed propagators in their evaluation, which would result in important
(but at present unknown) $\mathcal O(g^3)$ correction terms. For a discussion of this topic in the case of QCD, see Ref.~\cite{bn}.} Therefore, one must exercise caution in
interpreting these results and should preferably only use them when $\lambda\ll 10$. In Fig.~\ref{fig:pot}.b, we have circumvented this problem by plotting $D\lambda^2$
as a function of $m_\rmii{D}/T$, with $m_\rmii{S}^2/m_\rmii{D}^2$ fixed, but have now replaced it with an ambiguity related to the choice of
these ratios at any $\lambda\gtrsim 1$. From this figure it is, however, evident that the diffusion coefficient is relatively insensitive to deviations of the ratio
$m_\rmii{D}^2/m_\rmii{S}^2$ from the leading order weak coupling result of $\fr{1}{2}$.
\begin{FIGURE}[t]
{
\centerline{\epsfxsize=8.0cm\epsfysize=6.0cm\epsfbox{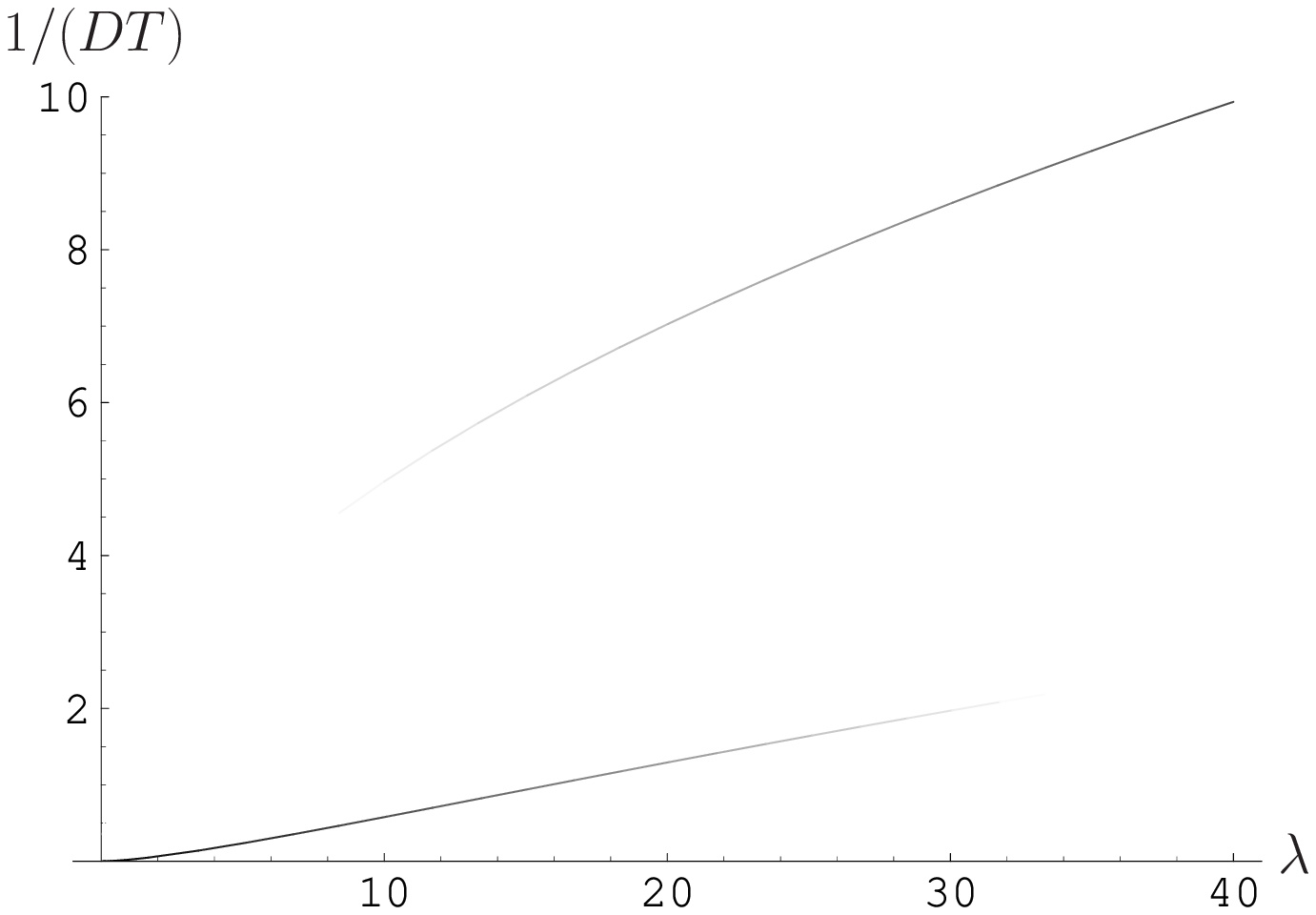}\epsfxsize=8.0cm\epsfysize=6.0cm\epsfbox{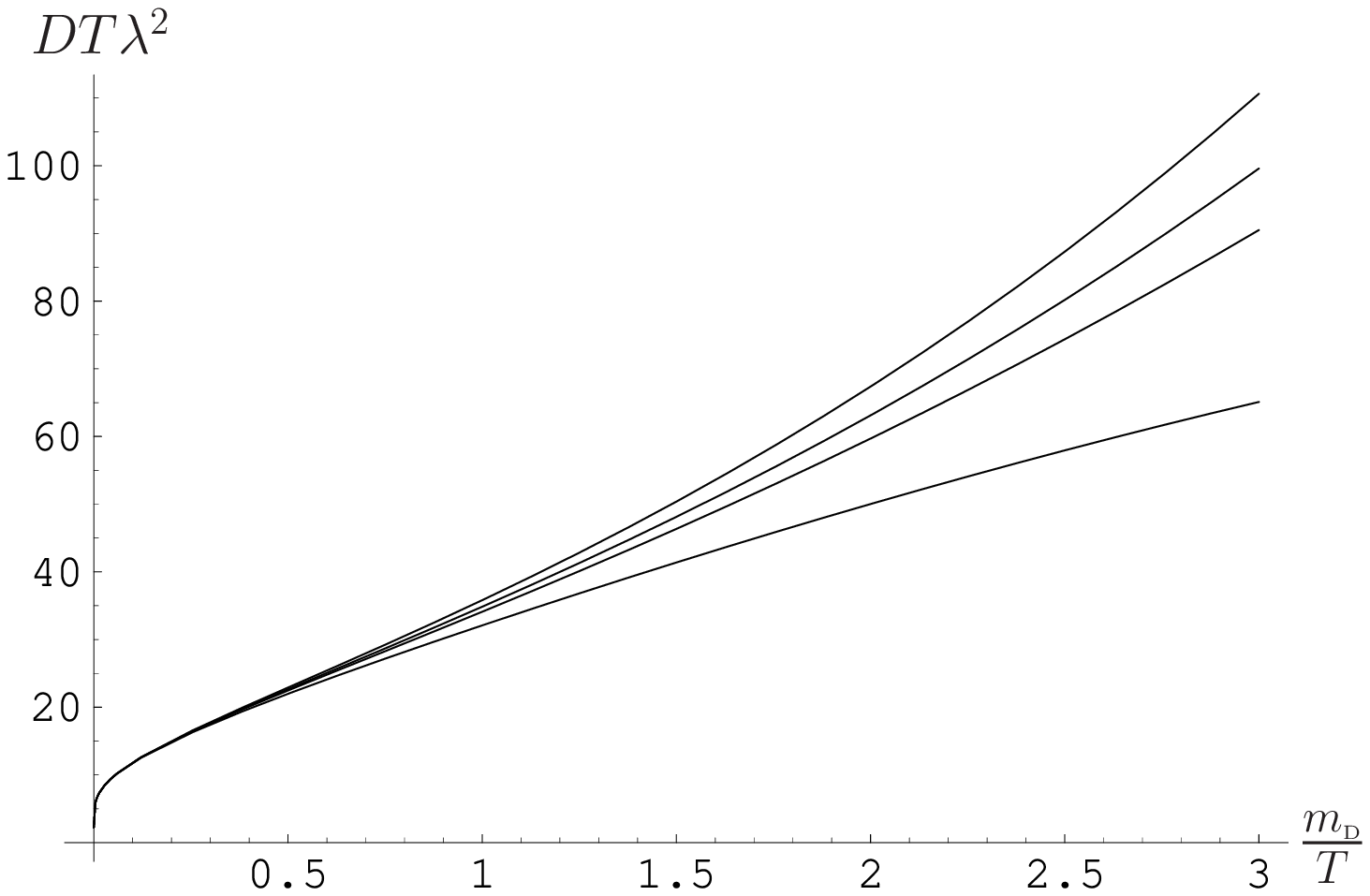}}
\caption[a]%
{Left: A plot of the weak (lower curve) and strong (upper curve) coupling results for $1/(DT)$ in the large $N_c$ limit. The weak coupling curve has been obtained via numerical
integration of Eq.~(\ref{fullkappaH}), with the screening masses given by their leading order perturbative expressions, while the strong coupling curve is taken from Eq.~(\ref{strong}).
\label{fig:pot} Right: The value of the (integrated) weak coupling result for $DT\lambda^2$ as a function of $m_\rmii{D}/T$ for $m_\rmii{S}^2/m_\rmii{D}^2=0$,
1/4, 1/2 and 1 (from bottom to top) in the large $N_c$ limit.}
}
\end{FIGURE}

For reasons of comparison, we have included already in Fig.~\ref{fig:pot}.a a plot of the strong coupling result for $1/(DT)$ as obtained from
Refs.~\cite{hky,ct}, according to which
\ba
\label{strong}
D = \frac{2}{\pi \sqrt{\lambda} T}
\ea
at large values of $\lambda$.\footnote{The region of validity of this result is currently somewhat unclear, but the authors
of Ref.~\cite{hky} point out that at the experimentally interesting couplings of $\lambda \sim 20$, it is already expected to obtain sizable corrections. Therefore,
we urge the reader to use this expression only for $\lambda\gg20$.}
As can be seen from Fig.~\ref{fig:pot}.a, it is easy to find a smooth, monotonic interpolating function that has the correct limiting behavior at small and large
$\lambda$, but at intermediate values of the coupling there is a wide region where neither result offers an accurate quantitative estimate for $D$.
At $\lambda \sim 20$, the weak coupling extrapolation of $D$ is seen to be roughly six times larger than the corresponding strong coupling prediction,
which is not surprising as we in any case are far beyond the region of validity of the weak coupling result here. However, we note that upon comparing
the forms of the weak and strong coupling curves at intermediate couplings, it appears
likely that the strong coupling expression yields an underestimate for the diffusion coefficient in this region.
\begin{FIGURE}[t]
{
\centerline{\epsfxsize=10.0cm\epsfysize=6.5cm\epsfbox{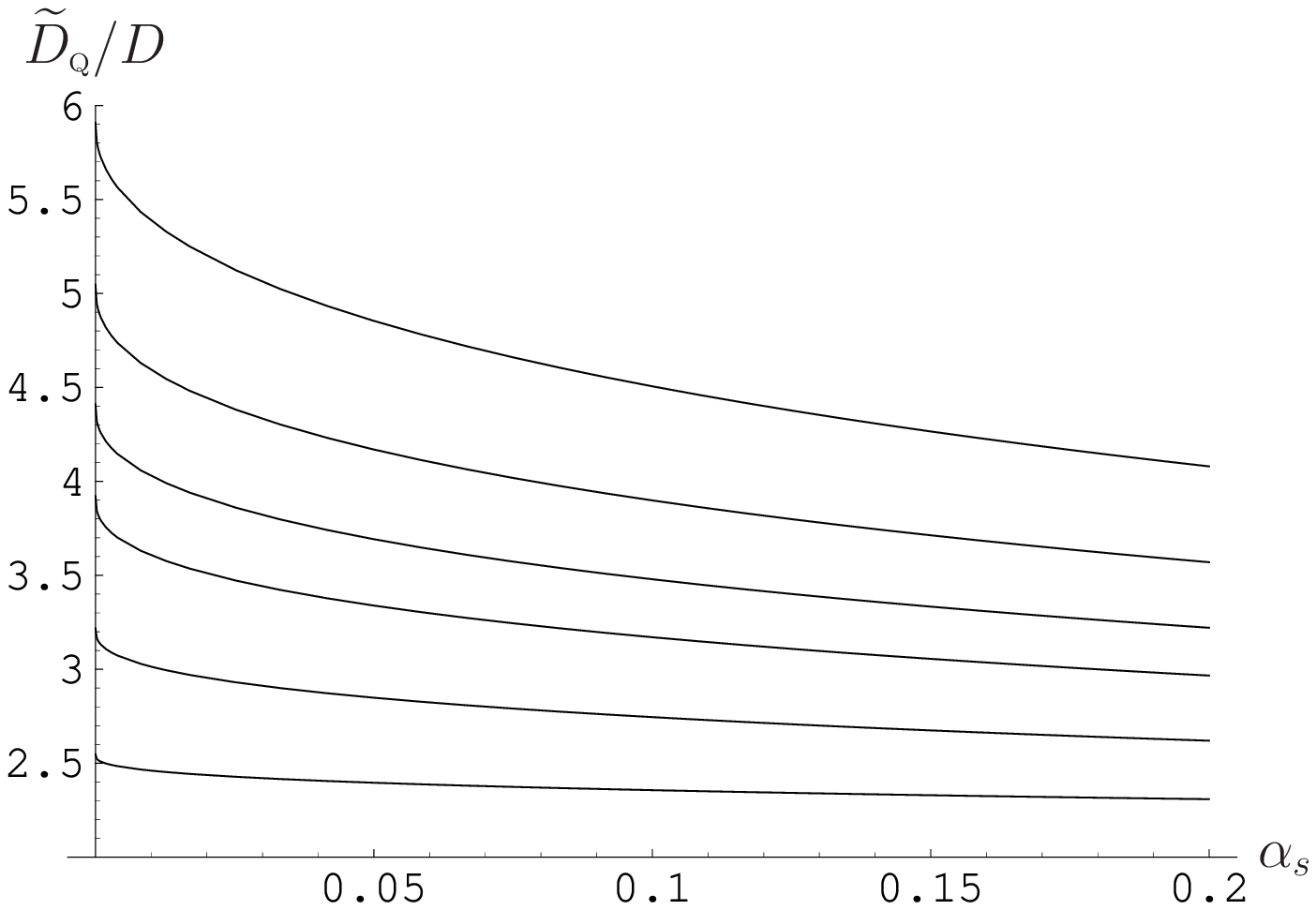}}
\caption[a]%
{Equal coupling plots of $\widetilde D_\rmi{Q}/D$ for $N_c=3$ and $N_f = 0, \, 1, \, 2, \, 3, \, 4$ and $8$ (from top to bottom)
as functions of $\alpha_s$.
}
\label{compare}
}
\end{FIGURE}

To put our discussion on a more quantitative footing, let us recall that using their weak coupling result, Moore and Teaney estimated the
heavy quark diffusion coefficient for $N_f=3$ QCD to be $\widetilde D_\rmi{Q} \approx 1/T$ at $\alpha_s= 0.5$, where we have adopted the convention of denoting QCD
quantities with tildes. In order to convert our result for the
$\mathcal N=4$ diffusion coefficient at $\alpha=0.5$ into at least a rough estimate for this quantity, we refer to the behavior of the ratio
$\widetilde D_\rmi{Q}/ D$ at weak (and equal) coupling, as shown in Fig.~\ref{compare}. There, $D$ is obtained by numerically integrating Eq.~(\ref{kappa}),
with the screening masses given by their weak coupling expressions at $N_c = 3$, while $\widetilde D_\rmi{Q}$ is evaluated from the equivalent QCD expressions \cite{mt}.
We see from this plot that at least at weak coupling, the ratio is a slowly decreasing function of $\alpha_s$,
and assuming this trend to carry on to stronger couplings, we estimate an upper bound
$\widetilde D_\rmi{Q}/ D \lesssim 3$ at $\alpha_s=\alpha =0.5$. Keeping in mind that the strong coupling result of Eq.~(\ref{strong}) is likely to be an overestimate at these couplings
(corresponding to $\lambda\approx 19$),
we on the other hand obtain $D|_{\alpha=0.5} \gtrsim 1/(7T)$, which translates
into the rough estimate $\widetilde D_\rmi{Q} \sim 3/(7T)\sim 1/(2T)$ at $\alpha_s= 0.5$.

Finally, from the point of view of the above comparisons between $\mathcal N=4$ SYM theory and QCD, it is also of some interest to investigate the ratio of the
diffusion coefficients of the two theories at very weak coupling to get some insight into the order of magnitude of this quantity.
Setting the couplings of the two theories again equal, we have from Ref.~\cite{mt}
\ba
\label{mote}
\widetilde D_\rmi{Q}&=& \fr{72\pi}{d_A g_s^4T}\bigg\{\log \fr{2T}{\widetilde{m}_\rmii{D}}+\fr{1}{2}-\gamma_E +\fr{\zeta'(2)}{\zeta(2)}\nn
&+& \frac{N_f}{2 N_c} \bigg ( \log \fr{4T}{\widetilde{m}_\rmii{D}}+\fr{1}{2}-\gamma_E +\fr{\zeta'(2)}{\zeta(2)} \bigg ) \bigg\}^{-1},
\ea
from which we obtain at asymptotically weak coupling (where the logs in Eqs.~(\ref{results}) and (\ref{mote}) dominate over the constant terms)
\ba
\fr{\widetilde D_\rmi{Q}}{D} \rightarrow \fr{6}{1+\fr{N_f}{2 N_c}}.
\ea
We observe that for all reasonable values of $N_f$, the QCD diffusion coefficient is considerably larger than that of $\mathcal N = 4$ SYM, which is mostly a reflection
of the fact that there are more light degrees of freedom in weakly coupled $\mathcal N = 4$ SYM theory for the heavy quark to scatter off of.
For example, in the $N_f=0$ case of pure Yang-Mills theory (in which all light particles are in the adjoint representation), one has 16 light bosonic
degrees of freedom, while $\mathcal N = 4$ SYM contains 64 bosonic and 64 fermionic degrees of freedom. A straightforward analysis shows that each
bosonic degree of freedom contributes equally to the leading $\log$ in $\kappa_\rmii{Q}$, while each fermionic degree of freedom contributes half as much, so that at asymptotically
weak coupling the diffusion coefficient of pure Yang-Mills theory should be $\fr{64+32}{16} = 6$ times bigger than that of $\mathcal N = 4$ SYM, just as we observed above.
Similar conclusions have been drawn also in Ref.~\cite{moorenew}, where the authors compared the weak coupling results for the sheer viscosity in $\mathcal N=4$ SYM and QCD.

\section{Conclusions and future directions}

In the paper at hand, we have investigated the diffusion of a heavy, non-relativistic thermal particle --- either a quark or a scalar belonging to a fundamental $\mathcal N=2$ hypermultiplet
--- immersed in $\mathcal N=4$ Super Yang-Mills plasma. We have derived a result for the heavy flavor diffusion coefficient that is valid to leading order in $g$ and $T/M$,
and compared it to the corresponding strong coupling results of Refs.~\cite{hky,ct} as well as to the weak coupling calculations of Ref.~\cite{mt} in QCD. Our findings show that
a naive extrapolation of the weak coupling result to intermediate couplings yields a relatively large disagreement with the strong coupling predictions, while in the weak coupling limit
the heavy flavor diffusion coefficient in the SYM theory is considerably smaller than the corresponding QCD quantity.
Based on our analysis, we have estimated the heavy quark diffusion coefficient in QCD to be roughly $\widetilde D_\rmi{Q} \sim 1/(2T)$ at $\alpha_s = 0.5$.

As is evident from the small number of weak coupling results available in $\mathcal N=4$ SYM theory, especially in comparison with the strong coupling limit or with
perturbative QCD, there is a lot of further work to be done that can provide the QCD community useful insights from the
abundance of existing AdS/CFT calculations. The obvious next goal related to the present work --- and one that that should be straightforward
to achieve --- is to generalize the results of this paper to the case of the diffusion of a relativistic quark with $\gamma v \gtrsim 1$. This is work in progress.

\section*{Acknowledgments}
We are grateful to Chris Herzog, Andreas Karch and Larry Yaffe for
suggesting the topic and for their valuable comments and advice, and
to Andy O'Bannon, Guy Moore, Maurizio Piai, Dam Son and Matt Strassler for
useful discussions. This work was supported by the U.S.~Department
of Energy under Grant No.~DE-FG02-96ER40956.

\appendix

\section{The Lagrangian}
\label{deriveL}
In this first Appendix, we aim to present a somewhat detailed derivation of the Lagrangian of our theory, $\mathcal N=4$ Super Yang-Mills with a massive $\mathcal N=2$
hypermultiplet, following to a large extent the treatment of Ref.~\cite{wein}.
The field content of the ${\mathcal N} =4$ theory consists of one gauge multiplet with components $\(A_{\mu}, \lambda, D\)$ and three chiral multiplets
$\chi$, $\chi'$ and $\chi''$ with components $\( \phi, \psi, \mathcal F\)$,  $\( \phi', \psi', \mathcal F'\)$ and $\( \phi'', \psi'', \mathcal F''\)$, respectively,
while the $\mathcal N=2$ sector contains two fundamental massive chiral multiplets $Q'$ and $Q''$ with components
$\(\Phi', \Psi', F'\)$ and $\(\Phi'', \Psi'', F''\)$. Here, $A_{\mu}$ is an $SU(N_c)$ gauge field, $\lambda$, $\psi$, $\psi'$, $\psi''$, $\Psi'$ and $\Psi''$ are
Majorana fermions, $\phi$, $\phi'$, $\phi''$, $\Phi'$, and $\Phi''$ are complex scalars, and $D$, $\mathcal F$, $\mathcal F'$, $\mathcal F''$, $F'$ and $F''$
so-called auxiliary fields. All fields in the $\mathcal N=4$ sector transform in the adjoint representation of the gauge group and are hence
traceless, hermitian $N_c\times N_c$ matrices, while the $\mathcal N=2$ fields are fundamental under SU($N_c$) and can therefore be viewed as $N_c$-component vectors
in color space.

If we fix the heavy particle masses to $M$, the superpotential of the theory reads
\ba
f(\chi,\chi',\chi'',Q',Q'')&=&-i \sqrt{2} \ Q''^{T}\chi Q'+2 i \sqrt{2} \ \tr(\chi[\chi',\chi''])+MQ'^{T} Q''.
\ea
Integrating out the auxiliary fields and going through some straightforward algebra, we obtain the Lagrangian \cite{wein}\footnote{In doing this,
we have identified and corrected several misprints in the original reference.
These are: an extra second term on the third-to last row of Eq.~(27.4.1), a missing $\e$ matrix between $\lambda$ and $\psi$ on the second row of Eq.~(27.9.3),
reversed indices $m$ and $n$ in the second term on the tenth row of Eq.~(27.9.33) and several misprints in the $\mu$-dependent terms of Eq.~(27.9.33). The
last two rows of the latter equation should be replaced by
\ba
&-& \frac{1}{4}(\mu^{\dagger} \mu)_{mn} \phi'^{\dagger}_m \phi'_n- \frac{1}{4}(\mu^{\dagger} \mu)_{mn} \phi''^{\dagger}_m \phi''_n
- \text{Re} \  \mu_{nm} \bar \psi'_n P_{+} \psi''_m\nn
&-&\sqrt{2} \ \text{Im} \ (t'_A \mu)_{mn} \phi_A \phi''^{*}_n \phi''_m -\sqrt{2} \ \text{Im}
\ (\mu t'_A)_{mn} \phi_A \phi'^{*}_m \phi'_n.
\ea}
\ba
\mathcal L &=& \mathcal L_0 + \mathcal L_1 + \mathcal L_2,
\ea
where
\vspace{-1.2cm}

{\setlength{\baselineskip}{1.6\baselineskip}

\ba
\label{l00}
\mathcal L_0&=&  -\tr \Big \{\frac{1}{2} F_{\mu \nu} F^{\mu \nu}+2 D_{\mu} \phi^{\dagger} D^{\mu} \phi + \bar \psi \slashed{D} \psi +  \bar \lambda \slashed{D} \lambda
+2 D_{\mu} \phi'^{\dagger} D^{\mu} \phi'
+2 D_{\mu} \phi''^{\dagger} D^{\mu} \phi''\nn
&+& \bar \psi' \slashed{D} \psi' +
\bar \psi'' \slashed{D} \psi''\Big \} -D_{\mu} \Phi'^{\dagger} D^{\mu} \Phi' - M^{2} \Phi'^{\dagger}\Phi'
-D_{\mu} \Phi''^{\dagger} D^{\mu} \Phi''  \nn
&-& M^{2} \Phi''^{\dagger} \Phi''- \frac{1}{2} \bar \Psi' \slashed{D} \Psi' -
\frac{1}{2} \bar \Psi'' \slashed{D} \Psi'' -2M \text{Re} \   \bar \Psi' P_{+} \Psi'',\\
\mathcal L_1/g &=& \text{Im} \Big \{-4 \sqrt{2} \tr  \bar \lambda P_{+} [\phi^{\dagger}, \psi] +4 \sqrt{2} \tr   \bar \psi' P_{+} [\phi, \psi'']
+4 \sqrt{2} \tr  \bar \psi'' P_{+}[\phi', \psi]
\label{l33} \nn
&-& 4 \sqrt{2} \tr  \bar \psi' P_{+}[\phi'', \psi]
+4 \sqrt{2} \tr \bar \psi' P_{+} [\phi'^{\dagger}, \lambda]
+4 \sqrt{2} \tr \bar \psi'' P_{+} [\phi''^{\dagger}, \lambda]  \nn
&-& 2 \sqrt{2} \ \bar \Psi'' P_{+} \psi \Phi' -2 \sqrt{2} \  \Phi''^{T} \bar \psi P_{+} \Psi' +2 \sqrt{2} \ \Phi'^{\dagger} \bar \lambda P_{+} \Psi'
- 2 \sqrt{2} \ \bar \Psi'' P_{+} \lambda \Phi''^{*} \nn
&-&2 \sqrt{2} M  \Phi''^{\dagger} \phi^{T} \Phi''
-2 \sqrt{2} M  \Phi'^{\dagger} \phi \Phi'  -2 \sqrt{2} \ \bar \Psi'' P_{+}\phi \Psi' \Big \},\\
\mathcal L_2/g^2 &=&-2 \tr \big | [\phi, \phi' ] \big |^2 -2 \tr \big | [\phi, \phi'^{\dagger} ] \big |^2-2 \tr \big | [\phi, \phi'' ] \big |^2
-2 \tr \big | [\phi, \phi''^{\dagger} ] \big |^2 \nn
&-&\fr{1}{2} \left | \tr t_a \big \{ 2 [\phi',\phi'^{\dagger}]
-2 [\phi''^{\dagger}, \phi''] + \Phi' \Phi'^{\dagger} - \Phi''^{*}  \Phi''^{T}\big \} \right |^{2} - \tr [\phi^{\dagger}, \phi]^{2} \nn
&-&2 \left | \tr t_a \big \{2 [\phi',\phi''] +\Phi' \Phi''^{T} \big \} \right |^{2}-\Phi'^{\dagger} \{ \phi,\phi^{\dagger} \} \Phi' -\Phi''^{T} \{ \phi,\phi^{\dagger} \} \Phi''^{*}.
\label{l44}
\ea
\par}
\noindent Here $P_{\pm}\equiv \fr{1}{2}(1\pm\gamma_5)$, $t_a$ are the generators of $SU(N_c)$ and a sum over $a$ is implied.

The form of the above functions can be greatly simplified upon making the redefinitions (adopted partially
from Ref.~\cite{yy})
\ba
\psi_1 &\equiv& \psi,\quad \psi_2 \equiv \lambda,\quad \psi_3 \equiv \psi', \quad \psi_4 \equiv \psi'', \nn
\omega &\equiv& P_{+} \Psi' + P_{-} \Psi'',\\
\phi_1 &=& \phi \;\, = \frac{1}{\sqrt{2}}(X_1 + i Y_1), \nn
\phi_2 &=& \phi' \, = \frac{1}{\sqrt{2}}(X_2+ i Y_2), \nn
\phi_3 &=& \phi''  = \frac{1}{\sqrt{2}}(X_3 + i Y_3), \\
\Phi_1 &\equiv& \Phi',\quad \Phi_2 \equiv \Phi''^{*},
\ea
where $X_p$ and $Y_p$ are hermitian scalar fields and $\omega$ a Dirac spinor. It is a straightforward exercise to show that in terms of these variables $\mathcal L_0$
reads

\vspace{-1.2cm}

{\setlength{\baselineskip}{1.6\baselineskip}
\ba
\mathcal L_0&=&  -\tr \Big \{\frac{1}{2} F_{\mu \nu} F^{\mu \nu}+ \bar \psi_i \slashed{D} \psi_i +\(D X_{p}\)^{2}+\(D Y_{p}\)^{2} \Big \}  \nn
&-&\Phi_n^{\dagger} (-D^{2}+M^2) \Phi_n - \bar \omega (\slashed{D} +M)\omega,
\ea
\par}

\noindent where a summation over repeated indices is implied. Using the Majorana condition of Eq.~(\ref{majorana}), the general form of the first six
terms in Eq.~(\ref{l33}) can on the other hand be written as

\vspace{-1.2cm}

{\setlength{\baselineskip}{1.6\baselineskip}
\begin{eqnarray}
4\sqrt{2} \,  \text{Im\,tr} \Big ( \bar \psi_i P_{+} [\phi_k,\psi_j] \Big )  &=& -i 2 \sqrt{2} \, \tr \Big ( \bar \psi_i P_{+} [\phi_{p},\psi_j] -
\bar \psi_i P_{-} [\psi_j, \phi^{\dagger}_{p}] \Big ) \nn
&=& - 2 \sqrt{2} \, \tr \Big ( i \bar \psi_i [\text{Re\,} \phi_p ,\psi_j] - \bar \psi_{i} \gamma_5 [\text{Im\,} \phi_{p}, \psi_{j}] \Big ),
\end{eqnarray}
\par}
\noindent which implies that one may simplify their sum into
\begin{eqnarray}
-\tr \Big \{i \bar \psi_i \alpha_{ij}^{p} [X_p,\psi_j] - \bar \psi_i \gamma_5 \beta^{p} _{ij}[Y_p, \psi_j] \Big \}.
\end{eqnarray}
Here, $\alpha^p$ and $\beta^p$ are coefficient matrices that may be taken as antisymmetric as the $\psi_i$ anticommute and whose components can easily be
verified to be given by Eq.~(\ref{coefficients}).  The remaining terms in Eq.~(\ref{l33}) are simple to translate into the new variables, leading to the final
result

\vspace{-1.2cm}

{\setlength{\baselineskip}{1.6\baselineskip}
\ba
\mathcal L_1/g &=& \tr \Big \{-i \bar \psi_i \alpha_{ij}^{p} [X_p,\psi_j] + \bar \psi_i \gamma_5 \beta^{p} _{ij}[Y_p, \psi_j] \Big \} -\bar \omega
 \left ( Y_1 - i \gamma_5 X_1 \right ) \omega  \nn
&+&2 \sqrt{2} \text{Im} \Big \{ - \bar \omega P_{+} \psi_1 \Phi_1 - \Phi_2^{\dagger} \bar \psi_1 P_{+} \omega +  \Phi_1^{\dagger} \bar \psi_2 P_{+} \omega
- \bar \omega P_{+} \psi_2 \Phi_2 \Big \} \nn
&-&2 M \Phi^{\dagger}_{n} Y_{1} \Phi_{n}.
\ea
\par}

Finally attacking $\mathcal L_2$, the terms in Eq.~(\ref{l44}) that are independent of $\Phi_n$ read

\vspace{-1.2cm}

{\setlength{\baselineskip}{1.6\baselineskip}
\ba
\label{noheavys}
&-&2 \tr \big | [\phi_1, \phi_2 ] \big |^2 -2 \tr \big | [\phi_1, \phi_2^{\dagger} ] \big |^2-2 \tr \big | [\phi_1, \phi_3 ] \big |^2
-2 \tr \big | [\phi_1, \phi_3^{\dagger} ] \big |^2 \nn
&-&\tr \left | [\phi_2,\phi_2^{\dagger}]
-[\phi_3^{\dagger}, \phi_3] \right |^{2} - \tr [\phi_1^{\dagger}, \phi_1]^{2} -4 \tr \left | [\phi_2,\phi_3] \right |^{2}.
\ea
\par}

\noindent Using the Jacobi identity for the cross terms on the second line, we have
\ba
-2 \, \tr [\phi_2,\phi^{\dagger}_2][\phi_3,\phi^{\dagger}_3] &=& 2 \,\tr \big | [\phi_2,\phi^{\dagger}_3] \big |^2 + 2  \, \tr   \big | [\phi_2,\phi_3] \big |^2,
\ea
so Eq.~(\ref{noheavys}) becomes

\vspace{-1.2cm}

{\setlength{\baselineskip}{1.6\baselineskip}
\ba
\label{noheavys2}
&-&2 \, \tr \big | [\phi_1, \phi_2 ] \big |^2 -2 \, \tr  \big | [\phi_1, \phi_2^{\dagger} ] \big |^2-2 \,  \tr \big | [\phi_1, \phi_3 ] \big |^2
-2 \, \tr \big | [\phi_1, \phi_3^{\dagger} ] \big |^2 \nn
&-&2 \,\tr \left | [\phi_2,\phi_3] \right |^{2} -2 \, \tr \left | [\phi_2,\phi^{\dagger}_3] \right |^2 - \tr [\phi_1^{\dagger}, \phi_1]^{2} -
\tr [\phi_2^{\dagger}, \phi_2]^{2}- \tr [\phi_3^{\dagger}, \phi_3]^{2} \nn
&=& -\frac{1}{2} \tr \( i [\chi_A, \chi_B]\)^2,
\ea
\par}
\noindent where we have defined $\chi_{A} \equiv (X_1,Y_1,X_2,Y_2,X_3,Y_3)$. As before, the remaining terms in Eq.~(\ref{l44}) are easy to translate
into the new variables, leading to the result

\vspace{-1.2cm}

{\setlength{\baselineskip}{1.6\baselineskip}
\ba
\mathcal L_2/g^2 &=&-\frac{1}{2} \tr \( i [\chi_A, \chi_B]\)^2+(-1)^{n}  \Phi_n^{\dagger} \( [\phi_2,\phi_2^{\dagger}] +
[\phi_3,\phi_3^{\dagger}] \) \Phi_n \nn
&-& 4 \text{Re} \( \Phi_1^{\dagger} [\phi_2,\phi_3] \Phi_2\)-\tfr{1}{2} \big | (-1)^{n} \Phi_{n}^{\dagger} t_a  \Phi_{n} \big |^{2}
-2  \big | \Phi_2^{\dagger} t_a \Phi_1 \big |^{2} \nn &-& \Phi^{\dagger}_{n} \{\phi_1,\phi_1^{\dagger} \} \Phi_{n},
\ea
\par}
\noindent where repeated indices are again summed over.

\section{Matrix elements and integrals}
\label{melemnts}

In this Appendix, we will briefly review our evaluation of the necessary scattering amplitudes squared in the non-relativistic limit, as well as explain, how
one can perform the integrals in Eq.~(\ref{fullkappaH}) analytically in the weak coupling limit. Our treatment is to a large extent parallel to that of Ref.~\cite{mt}.

\subsection{Matrix elements}
As discussed in Section \ref{calcs}, the scattering amplitudes squared for the heavy fermions and scalars become identical in the non-relativisic limit, which we
exploit by only computing the simpler scalar amplitudes in the Coulomb gauge. We denote the color, flavor and momentum of the initial and final light
particles by $a, m, \bv k$ and $b, n, \bv k'$, respectively, and the color and momentum of the heavy particles by $i, \bv p$ and $j, \bv p'$. The angle between $\bv k$ and $\bv k'$ is
written as $\theta$, the structure constants of the gauge group as $f_{abc}$ and the propagators for the scalars and the gauge field as $G^{p}_{cd}$ and
$D^{\mu \nu}_{cd}$. Because the plasma has no preferred color orientation, we adopt the convention of averaging over the color configurations of the initial heavy
particle, while the colors of the light particles as well as the final heavy scalar are summed over.

As a concrete example, consider the amplitude for the process $Sf \rightarrow Sf$. The total scattering amplitude for this process is the sum of the
first and fourth diagrams of Fig.~\ref{fig1}.b and is given by
\vspace{-1.2cm}

{\setlength{\baselineskip}{1.6\baselineskip}
\ba
\mathcal M_{Sf\rightarrow Sf} &=& \Big (-g\delta_{mn} f_{abc} \bar v(\bv k) \gamma_\mu v(\bv k') \Big ) D^{\mu \nu}_{cd}(Q) \Big ( i g(P + P')_{\nu}  (t_d)_{ij} \Big ) \nn
&+& \Big( g \beta^{1}_{mn} f_{abc} \bar v(\bv k') \gamma_5 v(\bv k) \Big) G^{1}_{cd}(Q) \Big (-2ig M (t_d)_{ij} \Big ).
\ea
\par}
\noindent Upon squaring this expression and summing over $m$ and $n$, it becomes evident that the cross term will be proportional to $\tr \beta^1$,
which vanishes due to the antisymmetricity of the matrix. Furthermore, in the non-relativistic limit we have $(P + P')_{\nu} \approx 2 M \delta_{\nu 0}$,
so after summing over the colors, flavors and spins of the light fermions as well as the colors of the final heavy scalar, we obtain the result of
Eq.~(\ref{scatter}a),
\ba
\left |\mathcal M_{Sf\rightarrow Sf} \right |^2 &=& 32  g^4 d_A M^2 k^2 (1+\cos \theta) \fr{1}{(q^2+m_\rmii{D}^2)^2} \nn
    &+&32 g^4 d_A M^2 k^2 (1-\cos \theta) \fr{1}{(q^2 + m_\rmii{S}^2)^2}.
\ea
The results quoted in Eqs.~(\ref{scatter}b)--(\ref{scatter}e) are obtained in a highly analogous fashion.

\subsection{Integrals}
The integrals appearing in Eq.~(\ref{fullkappaH}) are of the same type as those encountered in the QCD case, and our treatment of them therefore follows that
of Ref.~\cite{mt} quite closely. We begin by eliminating the three-dimensional delta function through integration over $\bv k'$, then change variables from $\bv p'$ to
$\bv q= \bv p'-\bv p$, and finally perform the angular part of the $\bv q$ integral to get rid of the energy delta function. This yields
\ba
\label{kappa}
3 \kappa_\rmii{H} &=& \fr{1}{64 \pi^3 M^2} \int_{0}^{\infty} dk \int_{0}^{2 k} dq\, q^3 \bigg \{ \fr{e^{\beta k}}{\(e^{\beta k} +1\)^2} \sum_{f, f'}
\left | \mathcal M_{Hf \rightarrow Hf'} \right |^2  \nn
&+& \fr{e^{\beta k}}{\(e^{\beta k} -1\)^2} \sum_{b, b'} \left | \mathcal M_{Hb \rightarrow Hb'} \right |^2 \bigg \},
\ea
with $\beta \equiv 1/T$. The sums over the amplitudes squared can be performed using the results of Eqs.~(\ref{scatter}a)--(\ref{scatter}e), and writing the results out
explicitly we get
\ba
\label{fermi}
\sum_{f, f'} \left | \mathcal M_{Hf \rightarrow Hf'} \right |^2 &=& 32 g^4 d_A M^2 k^2 \(2-\fr{q^2}{2 k^2} \) \fr{1}{(q^2+m_\rmii{D}^2)^2} \nn
&+& 32 g^4 d_A M^2 k^2 \(\fr{q^2}{2 k^2} \) \fr{1}{(q^2+m_\rmii{S}^2)^2},\\
\label{boson}
\sum_{b, b'} \left | \mathcal M_{Hb\rightarrow Hb'} \right |^2
&=&8 g^4 d_A  M^2 k^2 \( 8   -\fr{q^2}{k^2} + \fr {q^4}{4 k^4} \) \fr{1}{(q^2+m_\rmii{D}^2)^2} \nn
&+&16g^4 d_A M^2 k^2 \(\fr{q^2}{k^2} - \fr{q^4}{4 k^4} \) \fr{1}{(q^2+m_\rmii{S}^2)^2},
\ea
where the relation $\cos \theta = 1-q^2/2 k^2$ has been applied.

The integral over $q$ in Eq.~(\ref{kappa}) can be performed analytically, resulting in a somewhat lengthy one-dimensional integral representation for the diffusion
coefficient as a function of the screening masses, which we plotted numerically in Section 4. In the true weak coupling limit, where the screening masses satisfy
$m/T\sim g\ll 1$, we may, however, simplify the calculation considerably by noting that all terms in Eqs.~(\ref{fermi}) and (\ref{boson}) proportional to
positive powers of $q$ are infrared insensitive and thus independent of the masses to leading order. This enables us to set the masses to zero in these terms and
gives
\ba
\label{fermi2}
\sum_{f, f'} \left | \mathcal M_{Hf \rightarrow Hf'} \right |^2 &=& 64 g^4 d_A M^2 k^2 \fr{1}{(q^2 + m_\rmii{D}^2)^2},\\
\label{boson2}
\sum_{b, b'} \left | \mathcal M_{Hb\rightarrow Hb'} \right |^2 &=& 8g^4 d_A M^2 k^2 \( \fr{1}{q^2k^2} - \fr{1}{4k^4} + \fr{8}{(q^2+m_\rmii{D}^2)^2} \),
\ea
which simplifies the result of the $q$ integration in Eq.~(\ref{kappa}) dramatically. Finally evaluating the $k$ integrals with standard methods, our weak coupling
result for $3\kappa_\rmii{H}$ reads
\ba
\label{3kappafinal}
3\kappa_\rmii{H}&=&\fr{g^4 d_A}{2 \pi^3} \int_{0}^{\infty} dk k^2\bigg\{ \fr{e^{\beta k}}{\(e^{\beta k} +1\)^2} \(-1+\log\frac{4 k^2}{m_\rmii{D}^2} \)
+\fr{e^{\beta k}}{\(e^{\beta k} -1\)^2} \(-\fr{3}{4}+\log\frac{4 k^2}{m_\rmii{D}^2} \)\bigg\} \nn
&=&\frac{g^4 d_A T^3}{2 \pi} \bigg \{ \log \fr{2 T}{m_\rmii{D}} + \fr{13}{12} -\gamma_E + \frac{1}{3} \log 2 + \fr{\zeta'(2)}{\zeta(2)} \bigg\},
\ea
which in turn leads to the expression of Eq.~(\ref{results}) for $D_\rmi{Q}$ (or $D_\rmi{S}$).


\end{document}